\documentclass{jfm}
\usepackage{newtxtext}
\usepackage{newtxmath}
\usepackage{natbib}
\usepackage{hyperref}
\hypersetup{
    colorlinks = true,
    urlcolor   = blue,
    citecolor  = black,
    linkcolor  = black,
}

\newcommand{\RomanNumeralCaps}[1]
\linenumbers

\usepackage{graphicx,xcolor}
\usepackage{bm,soul}
\usepackage{physics}
\usepackage{comment}

\providecommand*{\rmd}{\mathrm{d}}
\renewcommand*{\epsilon}{\varepsilon}
\providecommand*{\pp}[3][]{\frac{\partial^{#1}#2}{\partial #3^{#1}}}

\providecommand*{\ppi[3][]}{\partial_{#3}^{#1}#2}
\providecommand*{\qq}[1]{\quad \mbox{#1} \quad}

\providecommand*{\rmd}{\mathrm{d}}

\newcommand{\BR}[1]{{\color{blue} #1}}

\newcommand{\lev}{\ell_{EV}}
\newcommand{\eit}{\mathrm{e}^{\mathrm{i} t}}

\renewcommand{\P}{\mathcal{P}}
\newcommand{\W}{\mathcal{W}}
\newcommand{\avg}[1]{\left<#1\right>}

\title{Elasto-inertial rectification of oscillatory flow in an elastic tube}

\author{Xirui Zhang\aff{1} and 
 Bhargav Rallabandi\aff{1} \corresp{\email{bhargav@engr.ucr.edu}}}

\affiliation{\aff{1}Department of Mechanical Engineering, University of California,
Riverside, CA 92521, USA}

\shorttitle{Elasto-inertial rectification of oscillatory flow in an elastic tube}
\shortauthor{X. Zhang and B. Rallabandi}

\begin{document}

\maketitle

\begin{abstract}
The interaction between deformable surfaces and oscillatory driving is known to yield complex secondary time-averaged flows due to inertial and elastic nonlinearities. Here, we revisit the problem of oscillatory flow in a cylindrical tube with a deformable wall, and analyze it under a long-wave theory for small deformations, but for arbitrary Womersley numbers. We find that the oscillatory pressure does not vary linearly along the length of a deformable channel, but instead decays exponentially with spatial oscillations. We show that this decay occurs over an elasto-visco-inertial length scale that depends on the material properties of the fluid and the elastic walls, the geometry of the system, and the frequency of the oscillatory flow, but is independent of the amplitude of deformation. Inertial and geometric nonlinearities associated with the elastic deformation of the channel drive a time-averaged secondary flow. We quantify this flow using numerical solutions of our perturbation theory, and gain insight into these solutions with analytic approximations. The theory identifies a complex non-monotonic dependence of the time-averaged flux on the elastic compliance and inertia, including a reversal of the flow. Finally, we show that our analytic theory is in excellent quantitative agreement with the three-dimensional direct numerical simulations of \citet{pande2023oscillatory}.  
\end{abstract}

\begin{keywords}
Authors should not enter keywords on the manuscript, as these must be chosen by the author during the online submission process and will then be added during the typesetting process (see http://journals.cambridge.org/data/\linebreak[3]relatedlink/jfm-\linebreak[3]keywords.pdf for the full list)
\end{keywords}

\section{Introduction}

The study of fluid flow in deformable geometries is important across various scientific domains, including physiology, biomedical engineering and lab-on-a-chip technologies \citep{das2010bio,yeh2017self, gervais2006flow,hardy2009deformation}. In laboratory settings, it is typical to fabricate fluidic systems with elastomeric materials \citep{ozsunNoninvasiveMeasurementPressure2013, ushayInterfacialFlowsArrays2023a}. The interplay between fluid flow and the mechanical response of channel walls excites nonlinear flow phenomena and has received renewed interest in recent years \citep{elbaz2014dynamics, christovFlowRatePressure2018,christovSoftHydraulicsNewtonian2022}. A number of these soft systems involve pulsatile, or more generally time-dependent flows \citep{kiranraj2019biomimetic, dincauPulsatileFlowMicrofluidic2020}, where a common feature is the rectification of time-periodic driving into steady ``streaming'' flows with nonzero time-average. This occurs due to a combination of inertial and geometric nonlinearities \citep{lafzi2020inertial,bhosaleSoftStreamingFlow2022,cuiThreedimensionalSoftStreaming2024}. Inertial (advective) nonlinearities drive streaming even in rigid systems \citep{gaverExperimentalInvestigationOscillating1986, riley2001steady,zhang2023three}. Conversely, geometric nonlinearities due to oscillatory elastic deformation can elicit a secondary steady response even without inertia \citep{zhang2020direct, kargar-estahbanatiLiftForcesThreedimensional2021,bureau2023lift}.

A configuration of particular interest involves time-periodic flows in compliant channels and conduits. Such flows occur naturally in physiological settings, e.g., in the flow of blood through both large arteries and small capillaries  \citep{pedley1980fluid, kiranraj2019biomimetic,baumler2020fluid,mirramezani2022distributed}, and the flow of air in the lungs. Oscillations may also occur spontaneously due to instabilities produced by large elastic deformations \citep{heil2003flows, grotbergBiofluidMechanicsFlexible2004,  rust2008feasibility, herradaGlobalStabilityAnalysis2022}. In engineered systems, interactions between oscillating flows and soft structures have found applications for flow control \citep{leslieFrequencyspecificFlowControl2009}, characterization of the dynamic properties of microchannel networks, \citep{vedelPulsatileMicrofluidicsAnalytical2010}, and in the design and control of valveless pumps with flexible membranes \citep{biviano2022smoothing,  amselemValvelessPumpingLow2023}. Furthermore, driving oscillations at high frequencies allows inertial effects to be exploited in compliant systems to design switches \citep{collino2013flow} and pumps \citep{zhangPowerfulAcoustogeometricStreaming2021}.

The earliest quantitative model of oscillatory flows in compliant tubes was due to \citet{womersleyOscillatoryMotionViscous1955} in the context of blood circulation. Womersley's work focused on the propagation of pressure pulses in long tubes and predicted a secondary flow due to deformation, but neglected the advective inertia of the fluid. Later work by \citet{dragonOscillatoryFlowMass1991} accounted for both inertial and geometric sources of streaming in a semi-analytic treatment, again focusing on wave-like pulse solutions in long tubes. Recently, \citet{pande2023oscillatory} developed a reduced-order model of pulsatile flow in compliant conduits of \emph{finite} length. Accounting for local fluid acceleration and elastic deformations, they obtained a nonlinear partial differential equation for the pressure distribution. The model yielded a secondary mean flow, which was found to be in agreement with three-dimensional (3D) direct numerical simulation (DNS) in stiff tubes, but underpredicted the pressure in more compliant tubes, where elastic deformations were greater.

Here, we revisit this problem, but take a different approach and develop a systematic perturbation theory for small elastic deformations in the vein of \citet{dragonOscillatoryFlowMass1991}. The analysis reveals an amplitude-independent elasto-viscous parameter, which plays an important role, along with the Womersley number (a measure of fluid inertia), in controlling the structure of the flow. By developing an approximate analytic solution of the theory, we obtain insight into both geometric and inertial mechanisms of driving time-averaged streaming flow over a wide range of elasto-viscous parameters and Womersley numbers. We find that both effects are equally important in understanding the time-averaged streaming flow, and their combination fully captures the aforementioned DNS results. 

The paper is organized as follows: Section \ref{secSetup} sets up the physical and mathematical problem, and identifies the important dimensionless groups that characterize the competition between viscous forces, fluid inertia, and elastic deformation. In Section \ref{secRes}, we develop a perturbation theory for small, but finite, deformations of the walls of the tube, and derive solutions for both the oscillatory and steady components of the flow. Analytic insight into the steady streaming is developed in Section \ref{SecTheory}, which we compare with the results of \citet{pande2023oscillatory} in Section \ref{secComp}. Finally, Section \ref{secSummary} concludes the paper by summarizing key findings and proposing future research directions.

\section{Model setup}\label{secSetup}
\begin{figure}
    \centering
    \includegraphics[width=\textwidth]{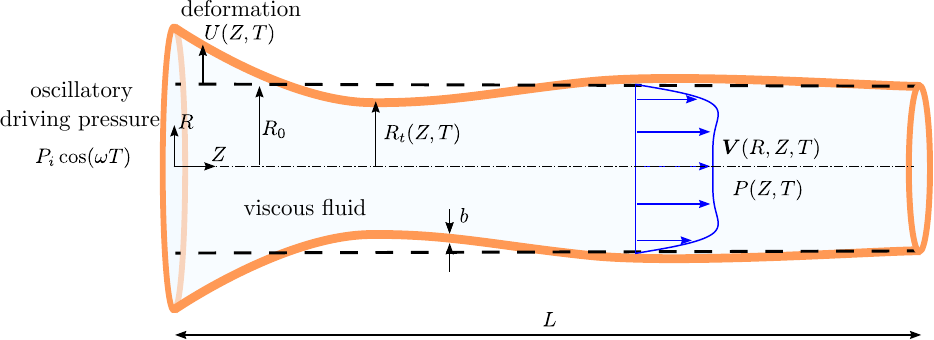}
    \caption{Schematic of setup:  A cylindrical tube with elastic walls is filled with a Newtonian fluid. Pressure oscillations applied at the inlet simultaneously drive fluid flow and deform the walls of the tube. The combination of boundary deformation and fluid inertia lead to a secondary flow with nonzero time-average.  }
    \label{3Dsetup}
\end{figure}
We consider the flow of Newtonian fluid with dynamic viscosity $\mu$ and density $\rho$ in a cylindrical tube of length $L$ and equilibrium radius $R_0$ (figure \ref{3Dsetup}). The walls of the tube have thickness $b$ and are made of an elastic material with Young's modulus $E$. An oscillatory pressure with amplitude $P_i$ and angular frequency $\omega$ is applied at the inlet of the channel ($Z=0$), relative to the outlet of the channel ($Z=L$) which is maintained at zero pressure. This sets up a fluid flow $\bm{V}(\bm{X},T)$ and an associated pressure field $P(\bm{X},T)$ in the tube, where $T$ represents time.  The stresses of the flow simultaneously deforms the walls of the tube. We denote the instantaneous radial displacement of the tube by $U(Z, T)$, so that the instantaneous radius of the tube is $R_t(Z, T) = R_0 + U(Z,T)$. 

Of particular interest is the time-averaged flow in the tube, which occurs both due to the elastic deformation of the tube, and the advective inertia of the flow. As we will see, both effects are equally important in this system, and both are activated by elasticity.

\subsection{Fluid flow}
We assume that the fluid flow is incompressible, axisymmetric and devoid of swirl. The flow velocity is denoted by $\bm{V}(Z,R,T) = \qty{V_R(Z,R,T), V_Z(Z,R,T)}$,  and satisfies the incompressible Navier-Stokes equations
\begin{subequations}\label{dimNS}
\begin{align}
    &\rho\left(\frac{\partial\bm{V}}{\partial T}+\bm{V}\cdot \nabla\bm{V}\right)=-\nabla P+\mu\nabla^2\bm{V}, \label{dimns} \\
    &\nabla\cdot\bm{V}=0 \label{dimcont}.
\end{align}
\end{subequations}
 It is useful to cast the continuity equation in terms of the flux $Q(Z, T) = 2 \pi \int_0^{R_t} V_Z R \rmd R$ by writing $\partial_Z Q + \partial_T (\pi R_t^2) = 0$, which leads to 
 \begin{align} \label{ContFluxDim}
     \pp{Q}{Z} + 2 \pi (R_0 + U)\pp{U}{T} = 0. 
 \end{align}
The flow is instantaneously no-slip at the walls  of the deforming tube ($\bm{V}= \ppi{R_t}{T} \bm{e}_R$ at $R = R_t$) and satisfies symmetry conditions at the centerline ($\ppi{V_Z}{R} = 0$, $V_R = 0$ at $R = 0$). The pressure oscillates as $P_i \cos (\omega T)$ at the inlet ($Z = 0$) and is zero at the outlet $(Z = L)$. 

\subsection{Elastic response of the tube}
To close the problem it is necessary to relate the deformation of the tube to the stresses in the flow. For thin-walled tubes ($b \ll R_0$), resistance to deformation occurs primarily due to elastic stresses generated by circumferential stretching of the tube \citep{landauTheoryElasticityVolume1986,audolyElasticityGeometryHair2018}. Focusing on slender channels, and anticipating the analysis of the subsequent sections, we approximate the normal stress exerted by the fluid on the tube by the fluid pressure $P(Z,T)$. For small deformations $U \ll R_0$, these arguments lead to the pressure-displacement relation 
\begin{align} \label{winkler}
    U(Z,T) = \frac{R_0^2}{E b} P(Z,T), 
\end{align}
which is both linear and local. This approximation of full three-dimensional elasticity is known to be accurate in similar systems \citep{rallabandi2021motion}. More generally, models of the kind in \eqref{winkler}, that relate local deformation to the local pressure (so-called Winkler foundation models), provide useful insight into flow-structure interactions in a wide range of systems \citep{ dillardReviewWinklerFoundation2018}. The finite bending rigidity and inertia of the walls of the tube have been neglected; their magnitude is estimated in Appendix \ref{ApxB}.

\subsection{Non-dimensionalization} \label{Nondim}
We rescale the equations by introducing dimensionless spatial coordinates (lowercase) $z=Z/ L$ and $r=R/ R_0$, and a dimensionless time $t=\omega T$. A balance between pressure gradients and viscous stresses sets a characteristic axial velocity scale $P_{i} R_{0}^{2}/(\mu L)$. By continuity, the radial velocity scale is $P_{i}R_{0}^{3}/(\mu L^{2})$. We  define a dimensionless pressure $p$, and dimensionless velocity components $v_z$ and $v_r$, according to
\begin{align} \label{NonDimVP}
    p =\frac{P}{P_i}, \quad v_z = \frac{V_z}{P_i R_0^2/(\mu L)},\qq{and} v_r=\frac{V_r}{P_i R_0^3/(\mu L^2)}.
\end{align}

The problem is governed by four independent dimensionless parameters, 
\begin{align}\label{dimlespara}
   \epsilon = \frac{R_0}{L}, \quad \W=\frac{ R_0^2  \omega}{\nu}, \quad\Lambda = \frac{P_i R_0 }{E b}, \qq{and} \sigma = \frac{\mu L^2 \omega }{E b R_0},
\end{align}
where $\nu = \mu/\rho$. Here, $\epsilon$ is the aspect ratio of the tube, and is small by construction. The Womersley number $\W$ is the ratio of a momentum diffusion timescale, $R_0^2/\nu$, to the driving timescale, $\omega^{-1}$.
The elasticity of the tube walls enters in \emph{two} distinct ways. The parameter $\Lambda$ characterizes the amplitude of radial deformation relative to the radius of the tube, and scales with the inlet pressure amplitude $P_i$. The linear elastic framework \eqref{winkler} implicitly assumes that $\Lambda$ is small, and we explicitly use this condition in the analysis of later sections. In addition to the deformation amplitude $\Lambda$, the flow-structure interaction introduces a timescale $\mu L^2/(E b R_0)$ over which elastic deformations relax against the viscous resistance of the fluid \citep[see, e.g.][]{elbaz2014dynamics}. The ratio of this relaxation time to the driving time defines an elasto-viscous parameter, $\sigma$, which is analogous to the Deborah number in rheology. Similar quantities arise in similar and other oscillatory fluid-elastic systems \citep{bickelHinderedMobilityParticle2007,leroyHydrodynamicInteractionsMeasurement2011,daddi-moussa-iderLonglivedAnomalousThermal2016,zhangContactlessRheologySoft2022,rallabandi2024fluid}. 

As we show in the next sections, the structure of the flow is governed primarily by $\W$ and $\sigma$, while $\Lambda$ sets the magnitude of the nonlinearity. We observe that $\W$ and $\sigma$ are similar in that they both represent the ratio of relaxation times to the driving timescale $\omega^{-1}$. We will later see that $\W$ and $\sigma$ also control the spatial structure of the flow in similar ways.

\section{Small-deformation theory for slender channels}\label{secRes}
We analyze the problem under a ``long-wave'' theory for slender channels ($\epsilon \ll 1$). The rescaled form of \eqref{dimNS} in this limit is \citep{dragonOscillatoryFlowMass1991,zhang2023three}
\begin{subequations}\label{NDgov}
\begin{align}
    &\W \qty{\frac{\partial v_z}{\partial t}+\frac{\Lambda}{\sigma} \qty(\bm{v}\cdot\nabla) v_z}=-\frac{\partial p}{\partial z}+\frac{1}{r}\frac{\partial}{\partial r}\left(r \frac{\partial v_z}{\partial r}\right) + O(\epsilon^2),\\
    &\frac{\partial p}{\partial r}=0 + O(\epsilon^2), \label{NDgovPressure}\\
    &\nabla \cdot \bm{v} = \frac{1}{r}\pp{(r v_r)}{r} + \pp{v_z}{z} = 0, 
\end{align}
\end{subequations}
where $\bm{v} \cdot \nabla = v_r \partial_r + v_z \partial_z$. Hereafter, we will not explicitly keep track of orders of $\epsilon$. In dimensionless variables, the wall of the tube is instantaneously at $r = 1 + \Lambda p(z,t)$. Rescaling \eqref{ContFluxDim} yields
\begin{align} \label{ContFlux}
    \pp{q}{z} + 2 \pi \sigma \pp{p}{t} (1 + \Lambda p) = 0, \qq{where} q(z,t) = 2 \pi\int_{0}^{1 + \Lambda p}  v_z r d r
\end{align}
is the instantaneous flux at any section of the channel, made dimensionless by $P_i R_0^4/(\mu L)$.  The axial flow satisfies no-slip at the walls of the tube,
\begin{align}
    v_z = 0 \qq{at} r = 1 + \Lambda p,
\end{align}
and the pressure field $p(z, t)$ satisfies the boundary conditions
\begin{align} \label{pBC}
    p\big|_{z= 0} = \cos t, \quad p\big|_{z= 1} = 0. 
\end{align}
It is interesting to note that the elasto-viscous parameter $\sigma$ appears both in the continuity equation and in the advective term.

We develop a perturbation theory for small, but nonzero, deformation amplitude $\Lambda \ll 1$, but make no restriction on $\mathcal{W}$ and $\sigma$. We seek a solution of the form 
\begin{subequations}\label{PertExp}
\begin{align} 
p(z, t) &= p_1(z, t) + \Lambda p_2(z, t) + O(\Lambda^2), \\
\bm{v}(r, z, t) &= \bm{v}_1(r, z, t) + \Lambda \bm{v}_2(r, z, t)  + O(\Lambda^2).
\end{align}
\end{subequations}
where, we have used \eqref{NDgovPressure} to eliminate the dependence of pressure on the radial coordinate. The subscript 1 identifies the primary oscillatory flow, which scales linearly with the pressure amplitude $P_i$ but is independent of the amplitude of elastic deformation. Secondary flow quantities (subscript 2) are quadratic in the pressure amplitude and are associated with the finite amplitude of elastic deformation. Substituting the series expansion \eqref{PertExp} into \eqref{NDgov} and separating orders of $\Lambda$ yields the governing equations for the primary and secondary flow, which we discuss in the next sections. %

It is useful to expand the flux and boundary conditions in powers of $\Lambda$. Substituting the expansion for $\bm{v}$ into \eqref{ContFlux}, the flux can be expanded as
\begin{subequations}\label{govq}
\begin{align}
    q(z,t) &= q_1(z,t) + \Lambda q_2(z,t) + O(\Lambda^2), \qq{where}\\
    q_1 &= 2\pi \int_0^1 v_{1z} r dr, \qq{and} q_2 = 2\pi \qty[\int_0^1 v_{2z} r dr + \qty{\qty(r v_{1z})  p_1}\Big|_{r =1}].
\end{align}
\end{subequations}
The second term in $q_2$ arises from the expansion of the limits of integration using Leibniz's theorem. We also Taylor expand the no-slip condition at the boundary wall $r=1+\Lambda p$ to obtain 
\begin{align} \label{NoSlipKappaExp}
    &v_{1z}+\Lambda \qty(v_{2z}+  p_1 \pp{v_{1z}}{r}) + O(\Lambda^2) =0 \qq{at} r=1.
\end{align}
We will use the above relations to analyze the primary and secondary contributions to the flow.  

\subsection{Primary flow} \label{SecPrimary}
At $O(\Lambda^0)$,  advective inertia  drops out of the momentum balance, yielding 
\begin{align}
    \W \frac{\partial v_{1z}}{\partial t}&=-\frac{\partial p_1}{\partial z}+\frac{1}{r}\frac{\partial}{\partial r}\left(r \frac{\partial v_{1z}}{\partial r}\right), \qq{with} 
    v_{1z}\Big|_{r = 1} = \pp{v_{1z}}{r}\bigg|_{r = 0} = 0. 
\end{align}
The continuity equation \eqref{ContFlux} at $O(\Lambda^0)$ is
\begin{align} \label{ContFluxPrim}
    \pp{q_1}{z} + 2 \pi \sigma \pp{p_1}{t} = 0. 
\end{align}

Motivated by boundary conditions for pressure, we seek solutions in terms of complex phasors oscillating as $e^{i t}$, whose real parts solve the physical problem.  
Writing the pressure as 
\begin{align}
    p_1(z, t) = \text{Re}[\mathcal{P}_1 (z) e^{i t}],
\end{align}
the oscillating axial velocity is found to be
\begin{align}
    &v_{1z}= \frac{1}{\alpha^2}\frac{\partial \mathcal{P}_1}{\partial z} \left(\frac{I_0(r \alpha )}{ I_0(\alpha ) }- 1\right) e^{i t},\qq{with}\alpha=\sqrt{i \W}, \label{v1xexpp13D}
\end{align}
where $I_n(\cdot)$ is a modified Bessel function of the first kind of order $n$. It will be understood that only the real part of complex oscillating quantities are physically meaningful. Equation \eqref{v1xexpp13D} is the well-known velocity profile identified by \citet{womersleyOscillatoryMotionViscous1955}, which takes on a parabolic shape for small $\W$, and a plug-like profile with boundary layers of dimensional thickness $R_0 \W^{-1/2} = (\nu/(R_0^2 \omega))^{1/2}$ for large $\W$. From \eqref{govq}, the primary oscillatory flux is found to be
\begin{align}
    &q_1=2\pi \int^1_0  v_{1z}\,r \text{d}r=-\frac{\pi I_2(\alpha)}{\alpha^2 I_0(\alpha)}\frac{\partial \P_1}{\partial z} \eit. \label{q1exp3D}
\end{align}

Substituting \eqref{q1exp3D} back into \eqref{ContFluxPrim} leads to an equation governing the complex pressure,
\begin{align}
    \frac{\partial^2 \P_1}{\partial z^2}-k^2 \P_1=0,\qq{where} k=\sqrt{\mathrm{i}\sigma \, \frac{2\alpha^2 I_0(\alpha) }{I_2(\alpha)}} \label{p1GE}
\end{align}
is a complex wavenumber. The above equation was also obtained by \citet{ramachandraraoOscillatoryFlowElastic1983} in analyzing linearized oscillatory flow in tapered elastic tubes and a variant of it is given in \citet{dragonOscillatoryFlowMass1991}. The primary pressure oscillates as $p_1= \cos t = \text{Re}[e^{it}]$ at the inlet $z =0$, and decreases to zero at the outlet ($p_1=0$ at $z=1$), yielding the boundary conditions $\P_1(0) = 1$, $\P_1(1)=0$. With these conditions, \eqref{p1GE} admits the solution
\begin{align}
        \P_1(z) =  \frac{\sinh\{k(1-z)\}}{\sinh{k}}\label{f13D}. 
\end{align}
Thus, we see that the primary pressure in a deformable tube decays as a spatially oscillating exponential. The limit $\sigma \to 0$ recovers the rigid case, where the pressure decreases linearly along the tube. 

\begin{figure}
    \centering
    \includegraphics[width=0.8\textwidth]{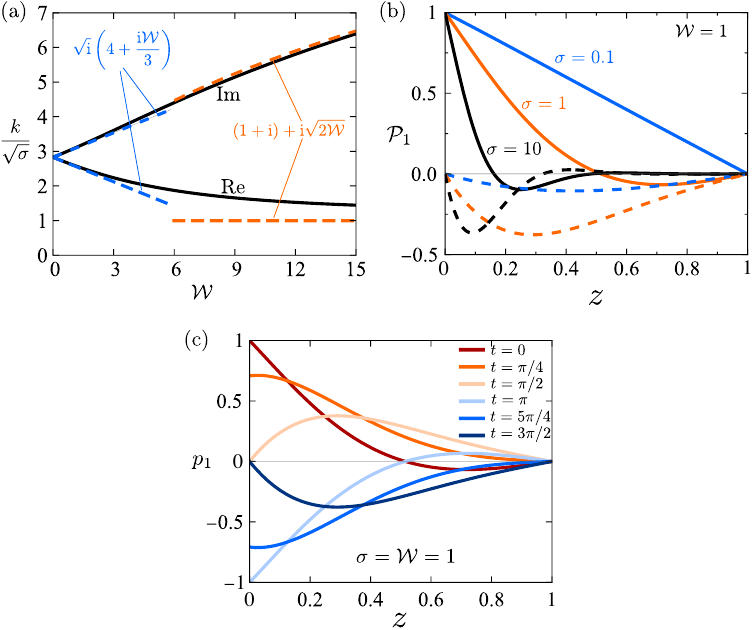}
    \caption{ (a) Complex wavenumber $k$ as a function of $\W$ showing real  and imaginary parts. Asymptotic expansions for large and small $\W$ are indicated by dashed curves. (b) Complex primary pressure $\P_1(z)$ at different $\sigma$ values. The pressure decays linearly along the tube for small $\sigma$, whereas it decays exponentially for larger $\sigma$. (c) Pressure distribution at different times during an oscillation cycle for $\sigma=\W=1$, indicating wave-like behavior.}
    \label{fig3Dp1}
\end{figure}

 The complex wavenumber  $k$ scales as $\sigma^{1/2}$ and depends on $\W$ through $\alpha$. Figure \ref{fig3Dp1}(a) shows the variation of $k/\sqrt{\sigma}$ with $\mathcal{W}$, indicating both real and imaginary parts. In the viscous limit $(\W \ll 1)$, $k$ approaches a constant $4 \sqrt{\mathrm{i} \sigma}$, indicating comparable rates of spatial decay and oscillation. For $\mathcal{W} \gg 1$, the real part of $k$ approaches $\sigma^{1/2}$, while the imaginary part diverges as $\sqrt{2 \sigma \mathcal{W}}$. Asymptotic behaviors for both small and large $\mathcal{W}$ are indicated in Figure \ref{fig3Dp1}(a).   Figure \ref{fig3Dp1}(b) shows both real and imaginary parts of $\P_1$ for different $\sigma$ at a fixed $\W$. For short relaxation times (small $\sigma$) the solid responds instantaneously to changes in the inlet pressure, and the pressure drops linearly in the tube, similar to a rigid channel. For longer relaxation times ($\sigma \gtrsim 1$), local deformation of the solid lags changes in inlet pressure, leading to a non-monotonic pressure distribution in the channel with comparable real (in-phase relative to the inlet) and imaginary (out-of-phase relative to the inlet) contributions.  At large $\sigma$ (and small $\W$) the pressure distribution decays axially from the inlet over a length scale $\lev = L \sigma^{-1/2} = \left(E b R_0 /(\mu \omega)\right)^{1/2}$. We observe that this is similar to how the Womersley velocity profile exhibits boundary layers of thickness $R_0 \W^{-1/2}$ for large $\W$. Thus, $\sigma$ controls its axial structure of the flow similarly to how $\W$ controls its radial structure. At larger values, $\W$ influences both radial and axial flow features; see \eqref{p1GE}, \eqref{f13D} and figure. \ref{fig3Dp1}. Figure \ref{fig3Dp1}(c) depicts profiles of  $p_1(z,t)$ at different times over an oscillation cycle for the particular combination $\sigma = \W = 1$. The pressure distribution travels along the channel in a wave-like pattern, gradually diminishing further away from the inlet of the tube. It is interesting to note that the elasticity of the tube strongly controls the oscillatory flow, even though the amplitude of deformation has not yet entered at this order in the analysis.

\subsection{Secondary flow} \label{SecSecondary}
With insight into the oscillatory flow structure, we study the flow at $O(\Lambda)$, associated with finite deformation of the channel walls. We focus specifically on the rectified (time-averaged) part of this flow, which is associated with a net flux through the channel. To this end, we time-average the momentum equation \eqref{NDgov} and the no-slip condition \eqref{NoSlipKappaExp} over an oscillation cycle. At $O(\Lambda)$, the averaged equations (using angle brackets to denote this average) are
\begin{subequations}\label{v2gov3D}
\begin{align}
        &\frac{\W}{\sigma}  \avg{\bm{v}_{1}\cdot\nabla v_{1z}}=-\frac{\partial \avg{p_2}}{\partial z}+\frac{1}{r}\frac{\partial }{\partial r}\left(r \frac{\partial \avg{v_{2z}}}{\partial r}\right), \qq{with} \label{v2gov} \\
        &\avg{v_{2z}} = v_{\rm slip}(z) \stackrel{\text{def.}}{=} - \avg{p_1 \pp{v_{1z}}{r}}\bigg|_{r =1} \qq{at} r = 1, \label{vslipdef} \\
        &\pp{}{r} \avg{v_{2z}} = 0 \qq{at} r = 0.
\end{align}
\end{subequations}
Here, $v_{\rm slip}(z)$ is an effective slip velocity satisfied by the secondary flow at the \emph{undeformed} location of the tube walls, and arises as a consequence of the domain perturbation in \eqref{NoSlipKappaExp}. A useful rule for computing the time-average of a product of oscillating complex quantities is $\left< \text{Re} \left[ A \mathrm{e}^{\mathrm{i} t}\right] \text{Re} \left[ B \mathrm{e}^{\mathrm{i} t}\right]\right> = (1/2) \text{Re}\left[A^* B\right] = (1/2) \text{Re}\left[A B^*\right]$; the asterisk denotes the complex conjugate \citep{longuet1998viscous}. We apply this rule to the definition \eqref{vslipdef} to calculate the effective slip velocity,
\begin{align}
        v_{\rm slip}(z)  = \frac{k I_1(\alpha)}{2\alpha I_0(\alpha)\qty|\sinh{k}|^2} \cosh\{k (1-z)\}\sinh\{k^*(1-z)\}, \label{reBC} 
\end{align}
where the real part is understood.

The problem \eqref{v2gov3D} is linear in $\avg{v_{2z}}$ and can be decomposed into three parts: (i) the velocity generated from advective ``body force'', which we denote $v_{2z}^\text{adv}$, (ii) the velocity associated with the geometric nonlinearity due to deformation, which manifests as the effective slip at $r =1$, and (iii) the velocity due to the secondary pressure gradient. Solving \eqref{v2gov3D} shows that the time-averaged axial velocity has the general structure
\begin{align} \label{v2zGeneral}
        \avg{v_{2z}}=\frac{1}{4}\frac{\partial \avg{p_2}}{\partial z}(r^2-1) + v_{\rm slip} + v_{2z}^\text{adv},
\end{align}
where the advective contribution $v_{2z}^\text{adv}$ satisfies
\begin{align} \label{AdvGE}
    \frac{\W}{\sigma}  \avg{\bm{v}_{1}\cdot\nabla v_{1z}}&=\frac{1}{r}\frac{\partial }{\partial r}\left(r \frac{\partial v_{2z}^\text{adv}}{\partial r}\right), \qq{with} 
    v_{2z}^{\rm adv}\Big|_{r = 1} = 0, \qq{and} \pp{v_{2z}^{\rm adv}}{r} \bigg|_{r = 0} = 0.
\end{align} 
The system \eqref{AdvGE} lacks a simple analytical solution due to the complicated radial dependence of the advective body force (involving products of Bessel functions), so we solve it numerically using finite difference techniques to find $v_{2z}^{\rm adv}$.

To obtain the secondary pressure, we first we insert \eqref{v2zGeneral} into \eqref{govq}, which yields a general expression for the mean flux
\begin{align}
        \avg{q_2} &= 2\pi \left(-\frac{1}{16}\frac{\partial p_2}{\partial z}+\frac{1}{2}v_{\rm slip} + \int_{0}^{1}v_{2z}^\text{adv}(r, z) r \mathrm{d}r\right). \label{3Dq2gov}
\end{align}
We then average the continuity equation \eqref{ContFlux} and collect terms at $O(\Lambda)$ to find that the time-averaged flux satisfies $\ppi{\avg{q_2}}{x} + 2 \pi \sigma \avg{\ppi{p_2}{t} + p_1 \ppi{p_1}{t} } = 0$. Observing that $\avg{\ppi{p_2}{t}} = 0$ and that $\avg{p_1 \ppi{p_1}{t} } = \avg{\ppi{}{t}\left(p_1^2/2\right)} = 0$, we find, unsurprisingly, that $\avg{q_2}$ is a constant. We use this result to integrate \eqref{3Dq2gov} along $z$, noting that $\avg{p_2}$ must vanish exactly at both ends of the tube, to obtain 
\begin{align} \label{q2General}
    \avg{q_2} = 2 \pi \int_0^1 \left(\frac{1}{2}v_{\rm slip}(z) + \int_{0}^{1}v_{2z}^\text{adv}(r, z) r \mathrm{d}r\right) \rmd z.
\end{align}
Rearranging \eqref{3Dq2gov}, the time-averaged pressure is therefore
\begin{align} \label{P2FormalSol}
    \avg{p_2}(z) &= 16 \int_0^z \qty(\frac{1}{2} v_{\rm slip}(x) + \int_{0}^{1}v_{2z}^\text{adv}(r, x) r \mathrm{d}r)dx - \frac{8 z}{\pi} \avg{q_2} 
\end{align}
Using \eqref{reBC} and the numerical solution to \eqref{AdvGE}, we implement the integrals in \eqref{q2General}--\eqref{P2FormalSol} numerically to obtain $\avg{p_2}(z)$. 

\begin{figure}
    \centering
    \includegraphics[width=\textwidth]{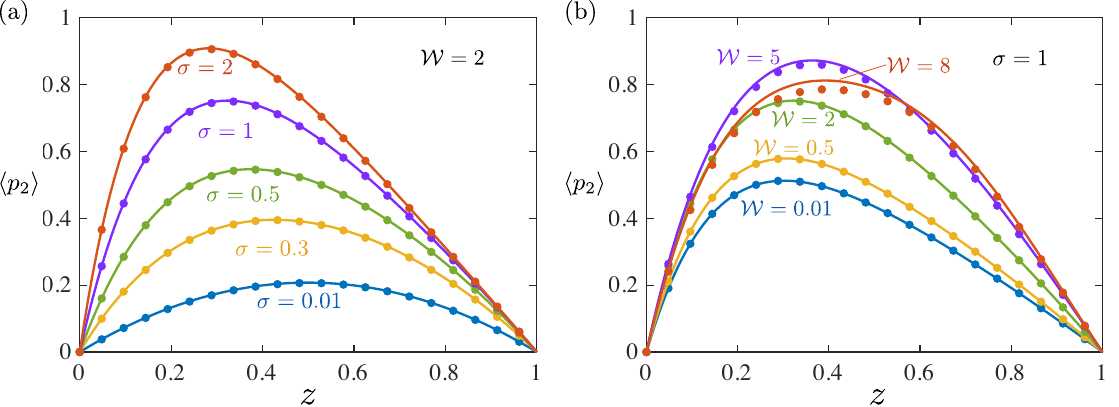}
    \caption{Secondary flow pressure distribution for (a) fixed  Womersley number $\W$ and (b) fixed elasto-viscous parameter $\sigma$. Symbols are results of numerical calculations and curves represent the approximate analytic theory. Panel (a) shows the transition from parabolic distribution in quasi-rigid channels ($\sigma \to 0$) to asymmetric patterns in flexible channels ($\sigma$ increases). Panel (b) demonstrates increasing peak pressure with rising $\W$ in flexible channels for a range of $\W$, though the dependence becomes non-monotonic for $\W \gtrsim 5$.}
    \label{fig3Dp2}
\end{figure}

The axial distribution of the secondary pressure is shown in figure~\ref{fig3Dp2} for different $\sigma$ and $\W$. Symbols represent the numerical solutions discussed above, and solid curves represent an analytic theory that we detail in section~\ref{SecTheory}. As $\sigma$ approaches zero for a fixed $\W$ (representing a nearly rigid channel with a short elasto-viscous relaxation time) the secondary pressure distribution becomes parabolic.  However, as $\sigma$ increases (slower elasto-viscous relaxation), the secondary pressure distribution becomes asymmetric, with the peak pressure rising and occurring closer to the channel inlet. For large $\sigma$, this peak occurs roughly at the elasto-viscous length scale $\lev$, but its precise location is set by the complex wavenumber $k$. Figure~\ref{fig3Dp2}(b) shows the dependence on $\W$ at a fixed value of $\sigma = 1$ (moderately flexible channels). As $\W$ increases, the peak of the secondary pressure increases up to $\W \approx 5$. For larger $\W$, the pressure begins to drop, deviating from the established pattern. Thus, the pressure distribution depends sensitively on both $\sigma$ and $\W$. Advective inertial effects become noticeable for $\W$ as small as $0.5$, and dominate for larger $\W$ (Appendix \ref{ApproxInertiaApp}; figure \ref{fig3Dslipcontri}).  These features  highlight the interplay between channel elasticity,  fluid inertia, and viscous resistance to flow, emphasizing the importance of both $\sigma$ and $\W$.

\section{Analytic approximation for the time-averaged flow} \label{SecTheory}
The complicated radial structure of the advective term in \eqref{AdvGE} precludes a simple analytic solution to $\avg{p_2}$. We gain analytic insight by approximating the advective term, by first constructing approximate oscillatory velocity fields with simpler radial dependences (Appendix \ref{ApproxInertiaApp}). Our approximation to the primary axial velocity, $\tilde{v}_{1z}$, shares the same centerline $(r = 0)$ velocity as the exact solution \eqref{v1xexpp13D}, but replaces the radial dependence with a parabolic no-slip profile. Then, we use the continuity equation to obtain an approximate primary radial velocity,  $\tilde{v}_{1r}$; see \eqref{v1app}. We then construct an advective term, $(\W/\sigma)  \avg{\tilde{\bm{v}}_{1}\cdot\nabla \tilde{v}_{1z}}$, which is given explicitly in \eqref{Aapp}. This approximation is asymptotic to the exact result in the viscous regime $\W \ll 1$, where the oscillatory axial velocity profile is indeed parabolic. Additionally, \eqref{Aapp} preserves the centerline value and the $z$-dependence of the exact advective term for \emph{all} $\W$, but presents a much simpler $r$-dependence that is now analytically tractable.

We use the result \eqref{Aapp} in place of the advective term in \eqref{AdvGE}, and solve the resulting equation to obtain an explicit analytic approximation to $v_{2z}^{\rm adv}$; see \eqref{v2advApp}. Substituting this result into \eqref{q2General}, we find that the mean flux is  
\begin{align} \label{q2theory}
    \avg{q_2}(\W, \sigma) &= \frac{\lambda \pi}{8}    \qty[ \qty(\frac{\cosh \{2 k_r\} - 1}{k_r}) +  \mathrm{i} \,\qty(\frac{ \cos \{2 k_i\} - 1}{k_i})], \qq{where} \\
    \lambda (\W, \sigma)&\simeq \underbrace{\frac{ k I_1(\alpha)}{\qty|\sinh{k}|^2\alpha I_0(\alpha) }}_{\text{effective slip}} + \underbrace{\frac{3 k^*}{32 \sigma} \qty| \frac{k}{\sinh{k}}\frac{I_0(\alpha)-1}{\alpha I_0(\alpha)}|^{2}}_{\rm advective\; inertia},
\end{align}
with $k_r = \Re{k}$ and $k_i = \Im{k}$.  The complex quantity $\lambda(\W, \sigma)$ is a scale factor comprising a contribution from the effective slip $v_{\rm slip}$ (which arises directly from the change in geometry due to deformation) and another from the advective inertia of the primary flow. We recall that $\alpha$ and $k$ depend on $\W$ and $\sigma$, see section \ref{SecPrimary}.
Using \eqref{P2FormalSol}, the approximate secondary pressure is 
\begin{align} \label{p2theory}
     p_2(z, \W, \sigma)  &=  \lambda(\W, \sigma)  \bigg[ \qty(\frac{z  + (1-z)\cosh\{2k_r\}-\cosh\{2k_r (1-z)\}}{k_r}) \nonumber \\
     &\qquad\qquad\;\; + \mathrm{i}\qty(\frac{z + (1-z)\cos\{2k_i\}-\cos\{2k_i (1-z)\}}{k_i})\bigg]. 
\end{align}
We note that the only source of approximation is in the advective term of $\lambda$; equations \eqref{q2theory}--\eqref{p2theory} are otherwise faithful reproductions of the exact solution to \eqref{v2gov3D}. Figure \ref{fig3Dp2} shows the theoretical pressure \eqref{p2theory} (solid curves) alongside ``exact'' (numerical) solutions (symbols). The theory is in excellent quantitative agreement with numerical calculations, up to $\W$ close to 10.

\begin{figure}
    \centering
    \includegraphics[width=1\textwidth]{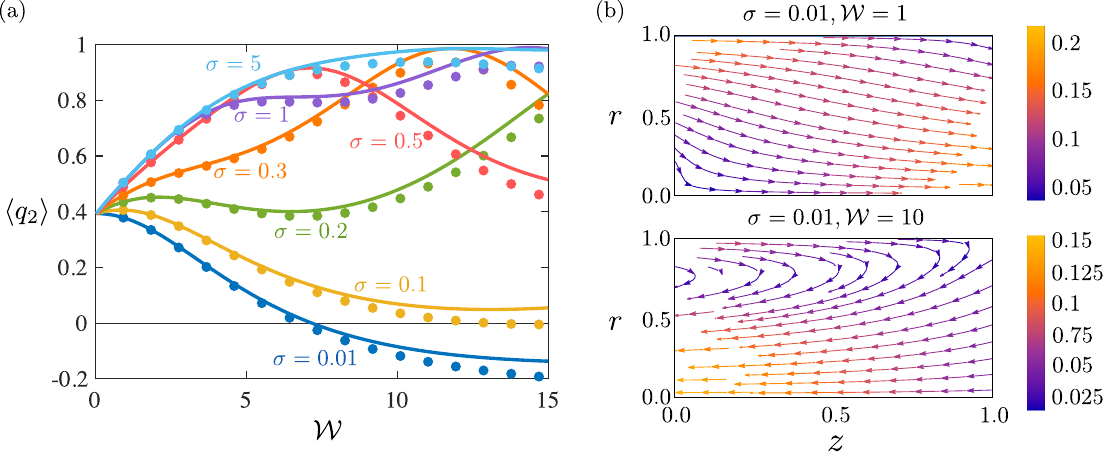}
    \caption{(a) Secondary flux for different $\sigma$ and $\W$.  In the case of small $\sigma$, the flux decreases as $\mathcal{W}$ increases, reaching negative values at large $\W$. As $\sigma$ increases, there is a general trend of increased secondary flux, which saturates at around 1 for large $\mathcal{W}$. (b) Example secondary flow visualization for $\sigma=0.01$ and $\W$: In a quasi-rigid channel, small $\W=1$ (Top) induces a secondary flow from inlet (the source of the oscillatory pressure) to outlet (held at zero pressure), while larger $\W$ leads to net flow from outlet to inlet. Color bars indicate dimensionless flow speed.}
    \label{fig3Dq2}
\end{figure}

The secondary flux is constant throughout the channel and is generally positive (i.e. towards the outlet), but varies for different values of $\sigma$ and $\W$, as shown in figure \ref{fig3Dq2}. The flux becomes independent of both $\W$ and $\sigma$ at small $\W$, converging to a value of $\pi/8$. When $\sigma$ is small (a relatively rigid channel) the flux decreases as $\W$ increases. It is striking that the flux eventually becomes \emph{negative} at sufficiently large $\W$ for small $\sigma$. However, for more flexible channels (larger $\sigma$), the secondary flux generally increases with $\W$, albeit non-monotonically. At larger values of $\mathcal{W}$, the flux saturates, and stays below unity. The dependence on $\sigma$ at fixed $\W$ is similarly complicated, displaying a non-monotonic behavior until saturating at large $\sigma$.   The theory \eqref{q2theory} (solid curves in figure \ref{fig3Dq2}) is in excellent agreement with numerical results (symbols) across all parameter values, and reproduces the generation of negative fluxes at small $\sigma$ and large $\W$, as well as the non-monotonic dependence on $\sigma$ and $\W$. The appearance of $k_i$ in the argument of a cosine in \eqref{q2theory} suggests that this non-monotonicity is due to the spatial oscillations of the primary pressure.

We now consider the limit $\sigma \ll 1$, where the elasto-viscous relaxation time is much smaller than the period of oscillation. In this limit, the walls of the tube respond quasi-statically to the oscillating flow, and the pressure decays linearly along the tube. In the limit, \eqref{q2theory} becomes
\begin{align} \label{q2rigid}
    \avg{q_2} \sim \frac{\pi}{8} \qty(\frac{2I_1(\alpha)}{\alpha I_0(\alpha) } - \frac{3 I_0(\alpha)}{8 I_2(\alpha)} \qty|\frac{I_0(\alpha)-1}{I_0(\alpha)}|^2) \qq{as} \sigma \to 0. 
\end{align} 
The first term is due to the slip velocity, while the second term is the contribution of inertia. It is interesting to note that advective inertia survives even in the quasi-rigid limit $\sigma \to 0$ (even as $\bm{v} \cdot \nabla \bm{v}$ becomes vanishingly small) due to the appearance of a $\sigma^{-1}$ in front of the advective term in \eqref{v2gov3D}. Plotting each term separately shows that the contribution of slip is always positive, whereas the contribution of inertia is always negative for $\sigma \to 0$. The inertial term is small for small $\W$, leading to a positive, slip-dominated, mean flow. At larger $\W$, the contribution of inertia to the flux outweighs that of the slip, leading to a negative mean flow. From \eqref{q2rigid}, we find that the transition from positive to negative flux occurs at $\W \approx 6.9$, consistent with the $\sigma = 0.01$ curve in figure \ref{fig3Dq2}. Streamlines of the steady flow across this transition are plotted in figure \ref{fig3Dq2}, showing that the flow near the wall $r = 1$ (due to the effective slip) is always to the right even as the net flow changes sign with $\W$. A similar analysis of \eqref{p2theory} shows that the secondary pressure has the form 
\begin{align}
    \avg{p_2} \sim \frac{8 \avg{q_2}}{\pi}\, z (1-z) \qq{for} \sigma \to 0.
\end{align}
The reversal in flux is thus accompanied by a reversal in the pressure. For small $\W$ (in addition to small $\sigma$), we find
\begin{align} \label{q20}
    \avg{q_2} \sim \frac{\pi}{8} \qty( 1 - \frac{5 \W^{2}}{96} + O(\W^3) ) \qq{for} \sigma \to 0.
\end{align}
The leading term is consistent with past work \citet{pande2023oscillatory, pande2023reciprocal}, while the second term differs due to the inclusion of advective inertia here. 

These behaviors change sensitively as $\sigma$ increases even slightly from zero. Both the slip and advective contributions become more positive with growing $\sigma$. At $\sigma \approx 0.1$, the reversal of flux with $\W$ no longer occurs (figure \ref{fig3Dq2}a). For $\sigma \gtrsim 0.2$, the advective term becomes positive at all $\W$, thus aiding the slip velocity. As a result, the flux grows with $\W$ when $\sigma \gtrsim 0.2$, in contrast to the $\sigma \ll 1$ case (figure \ref{fig3Dq2}). The flux eventually saturates for large $\sigma$, once again becoming independent of it. In the limit, and for small $\W$, \eqref{q2theory} yields
\begin{align}
    \avg{q_2} \sim \frac{\pi}{8} \qty( 1 + \frac{5 \W}{16} + O(\W^2)), \qq{as}{\sigma \to \infty}. 
\end{align}
We observe that the flux increases linearly with $\W$ at large $\sigma$, in contrast to the quadratic decrease at small $\sigma$ in \eqref{q20}. The contribution of inertia to the secondary flux becomes noticeable for $\W \gtrsim 0.5$, and may dominate for larger $\W$; see figure \ref{fig3Dslipcontri}. 

\section{Comparison with the simulations of \citet{pande2023oscillatory}}\label{secComp}

\begin{figure}
    \centering
    \includegraphics[width=1\textwidth]{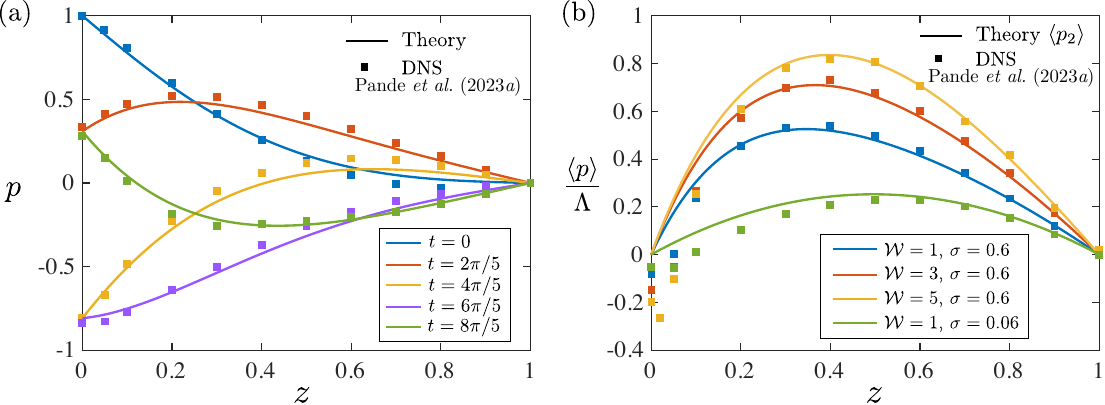}
    \caption{Comparison between the present analytic theory (curves) with the 3D DNS of \citet{pande2023oscillatory} (symbols), with  $\varepsilon = 0.0667$ and $\Lambda = 0.05$. (a) Instantaneous pressure distribution over an oscillation cycle for $\sigma=0.6$ and $\W=1$, and (b) Time-averaged pressure for different $\sigma$ and $\W$.}
    \label{fig3dcompar}
\end{figure}

\citet{pande2023oscillatory} conducted three-dimensional (3D) direct numerical simulations (DNS) of the problem, utilizing the svFSI finite-element solver of the open-source cardiovascular modeling software \emph{SimVascular} \citep{updegrove2017simvascular,lan2018re}. These simulations model the solid as a Kirchhoff--Saint Venant hyperelastic material, and fully resolve the two-way coupled time-dependent fluid-structure interaction \citep{zhu2022svfsi}.

Figure \ref{fig3dcompar}(a) shows the pressure distribution $p(z,t)$ at different times over an oscillation cycle, comparing DNS (square symbols) and our analytic theory (curves), which use the approximation $p(z,t) \approx  p_1(z,t) + \Lambda \avg{p_2}$. We note that in the theory we do not resolve the time-dependent part of $p_2$, though the error from this omission to the instantaneous pressure (which is dominated by $p_1$) is small. The agreement is excellent for $\sigma=0.6$ and $\W=1$, with no adjustable parameters. There is no discernible difference between our approximate analytical solutions and our numerical solutions for these parameters.

Figure \ref{fig3dcompar}(b) shows the time-averaged part of the pressure $\avg{p}/\Lambda = \avg{p_2}$, comparing again the DNS (symbols) and our theoretical result \eqref{p2theory} (curves). For different combinations of $\W$ and $\sigma$, the analytic theory captures the DNS remarkably well, again with no adjustable parameters. Our numerical calculations and analytic theory (cf. section \ref{SecSecondary}) are identical to within the thickness of the curves in figure \ref{fig3dcompar}(b). 

In addition to performing DNS, \citet{pande2023oscillatory} developed a reduced-order model that accounted for the local acceleration of the fluid but neglected its advective inertia. That model, when solved numerically, recovered the DNS results for $\W \lesssim 1$ with  $\sigma = 0.06$, but deviated for larger $\W$ and $\sigma$. The agreement of our theory with the DNS shows that these deviations can be almost entirely understood by accounting for the advective inertia of the oscillatory flow. While additional geometric or material nonlinearities associated with the response of the elastic solid may indeed be important for larger deformations, they do not appear to play a significant role for $\Lambda = 0.05$. It remains to be seen how far the theory can be pushed before departures from DNS become appreciable.

\section{Conclusions}\label{secSummary}

We have shown how the combination of fluid inertia and elastic deformations rectifies oscillatory driving in compliant channels into steady secondary flows. We quantify these features using a long-wave perturbation theory for slender channels and small elastic deformations. The amplitude of deformation governs the magnitude of nonlinearities driving steady flow. The structure of the flow, however, is independent of this amplitude and depends instead on two independent parameters: the Womersley number $R_0^2 \omega/\nu$, which measures a momentum diffusion timescale against the timescale of driving, and an elasto-viscous parameter $\mu L^2 \omega/(E b R_0)$, which is the ratio of an elasto-viscous relaxation time to the driving timescale. The precise combination of these parameters controls the spatial structure of the oscillatory flow, which in turn determines the relative importance of geometric nonlinearities (due to deformation) and advective nonlinearities (due to fluid inertia) in driving time-averaged flow. Using judicious approximations for the advective inertia, we obtained closed-form analytic approximations that quantitatively recover 3D DNS across the entire parameter space. 

The insights derived from the present work enable systematic understanding and efficient quantification of time-dependent fluid-structure interactions in practically relevant settings. Future work may account for the effect of a finite bending rigidity and the inertia of the elastic material. It would be equally interesting to quantify the effects of internal dissipation in the solid, which would be relevant in practical systems, particularly at large frequencies \citep[see, e.g., ][]{anand2020transient}. We expect that the  small-deformation framework developed here could be expanded to accommodate these features in order to gain substantial efficiencies over direct numerical simulations. The ideas developed here thus hold the potential to influence advancements in  engineered and naturally occurring systems such as cardiovascular physiology, biomedical engineering, and microfluidic device design.

\backsection[Acknowledgements]{
We thank I. C. Christov and S. Pande for motivating this work through a talk at an American Physical Society Division of Fluid Dynamics meeting, and for helpful discussions.}

\backsection[Funding]{ The authors thank the NSF for support through awards CBET-2143943 and CBET-2328628. B. R. acknowledges partial support through NSF award CBET-2126465. X. Z. acknowledges partial support through a dissertation year fellowship from the University of California, Riverside. }

\backsection[Declaration of interests]{The authors report no conflict of interest.}

\backsection[Author ORCIDs]{X. Zhang, https://orcid.org/0000-0002-0880-3911; B. Rallabandi, https://orcid.org/0000-0002-7733-8742. }

\appendix

\section{Bending rigidity and solid inertia} \label{ApxB}
We discuss two potential sources of deviation from the local elastic model \eqref{winkler}. 

\subsection{Finite bending rigidity}
Accounting for a finite bending rigidity $B \propto E b^3$ adds a term $\propto R_0^2 B/(E b) \ppi[4]{U}{Z} \propto b^2 R_0^2 \ppi[4]{U}{Z}$ (in addition to sub-dominant contributions) to \eqref{winkler} \citep{landauTheoryElasticityVolume1986}. Following the arguments of section \ref{SecPrimary}, the ratio of bending to stretching stresses is then $b^2 R_0^2 |k|^4$, with $k$ given by \eqref{p1GE}. We find that for small $\W$, bending becomes important when $\sigma \gtrsim L^2/(R_0 b)$, which is very large by definition. For large $\W$, the present theory identifies spatial oscillations at a wavenumber $\sqrt{\sigma \W}$ (Fig. \ref{fig3Dp1}). Bending rigidity of the tube suppresses these oscillations when $\sigma \gtrsim  L^2/(R_0 b \W^4)$. For small $\W$, bending only becomes important at very large $\sigma$. However, for modestly large inertial effects, particularly when $\W$ is comparable with $(L^2/(R_0 b))^{1/4}$, these effects may become noticeable for moderate $\sigma$. For example, with $L_0 \approx 1$ cm, $R_0 \approx 1$ mm and $b \approx 0.1$ mm, bending effects may start to play a role for $\sigma \simeq 1$ and $\W \simeq 6$. However, the favorable agreement of the present theory with DNS for a similar combination of parameters (figure \ref{fig3dcompar}) seems to indicate that the hoop-stress-only model may be more robust than is suggested by this scaling estimate.

\subsection{Solid inertia}
The inertia of the walls of the tube becomes more important at large $\W$,  and scales (per area) $\rho_s b \ppi[2]{U}{T}$, where $\rho_s$ is the density of the elastic solid. It becomes comparable to the elastic hoop stress $E b U /R_0^2 $ when $\omega \gtrsim \left(E/(\rho_s R_0^2)\right)^{1/2}$, i.e. $\W  \gtrsim \left(E R_0^2/(\rho_s \nu^2)\right)^{1/2}$. For a tube of radius $1$ mm and a typical soft material ($E \approx 1$ MPa) carrying water, this occurs for $\W \gtrsim 1000$. This is much greater than the $\W$ considered here, so solid inertia is expected to play a negligible in the present study.

\section{The role of advective inertia}\label{ApproxInertiaApp}
\subsection{Analytic approximation of inertial effects}
To approximate the advective inertia of the primary flow, we start by constructing approximations to the primary  velocity. The approximate axial velocity is constructed to share the same centerline value as its exact counterpart \eqref{v1xexpp13D}. Continuity determines the corresponding (approximate) radial velocity. This results in
\begin{align} \label{v1app}   
    \tilde{\bm{v}}_1 = \qty{\tilde{v}_{1r},\; \tilde{v}_{1z}} &=  \frac{1}{\alpha^2} \left(\frac{I_0(\alpha )-1}{ I_0(\alpha ) }\right) \qty{  \pp[2]{\P_1}{z} \qty(\frac{r}{2} - \frac{r^3}{4})e^{i t},\; -\pp{\P_1}{z} (1 - r^2) e^{i t}}. 
\end{align} 
The approximate advective body force due to the primary flow is (writing $\W = -i \alpha^2$)
\begin{align} 
        \frac{\W}{\sigma}\left<{\bm{v}}_{1}\cdot\nabla {v}_{1z}\right> \approx \frac{\W}{\sigma}\left<\tilde{\bm{v}}_{1}\cdot\nabla \tilde{v}_{1z}\right> &= -\frac{k^*}{4 \sigma}{\bigg| \frac{k}{\sinh{k}}\frac{I_0(\alpha)-1}{\alpha I_0(\alpha)} \bigg| }^2  (r^4-2 r^2+2) \nonumber \\
        &\qquad \qquad  \cosh\{k(1-z)\}\sinh\{k^*(1-z)\}. \label{Aapp}
\end{align}
Substituting \eqref{Aapp} into \eqref{AdvGE} and solving leads to 
\begin{align} 
    v_{2z}^{\rm adv} &\simeq   -\frac{k^*}{288\sigma} {\bigg| \frac{k}{\sinh{k}}\frac{I_0(\alpha)-1}{\alpha I_0(\alpha)} \bigg| }^2 (2 r^6-9 r^4+36 r^2-29) \nonumber \\
    &\qquad \qquad \qquad \qquad \cosh \{k(1-z)\}\sinh\{k^*(1-z)\}. \label{v2advApp}
\end{align}

\begin{figure}
    \centering
    \includegraphics[width=\textwidth]{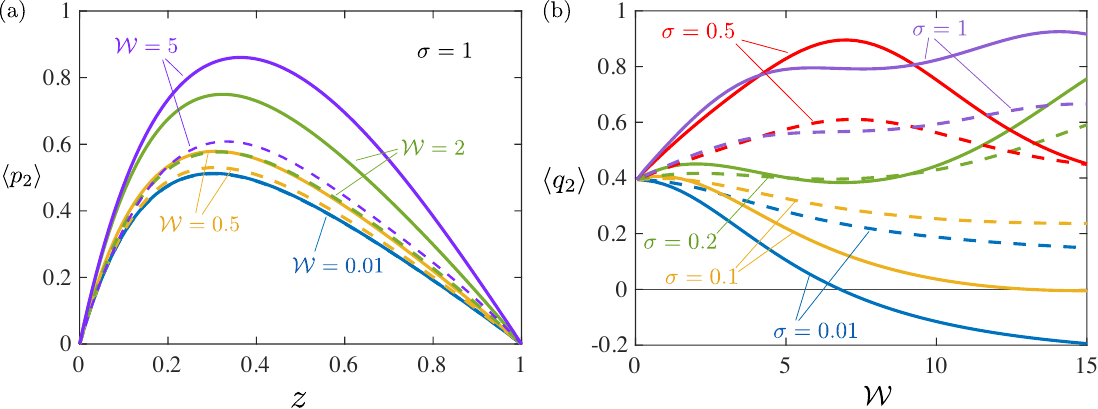}
    \caption{Contribution of advective inertia to (a) the secondary pressure (a) and (b) the secondary flux. Solid curves represent the full theory, while the dashed curves depict the theory in the absence of advective inertia.}
    \label{fig3Dslipcontri}
\end{figure}

\subsection{Relative magnitude of inertial and geometric nonlinearities}
To assess the contributions of advective inertia to the secondary flow, we solve a version of \eqref{v2gov3D} that neglects the advective term. This results in secondary pressures and fluxes that depend on the geometric nonlinearity only, which enters through the effective slip $v_{\rm slip}$. Figure \ref{fig3Dslipcontri} shows secondary pressures and fluxes computed in this way, both  with inertia (solid curves), and without inertia (dashed curves). The contributions of advective inertia are important for $\W \gtrsim O(1)$.

\end{document}